\documentclass[aps,prd,preprint,superscriptaddress,tightenlines,nofootinbib,showpacs,floatfix]{revtex4}

\usepackage{graphicx}
\usepackage{dcolumn}
\usepackage{bm}

\begin{document}

{\vbox{\hbox{\hfil CLNS 08/2026}\hbox{\hfil CLEO 08-09}}}

\def\Rg{$R_\gamma=0.137\pm0.001\pm0.016\pm0.004$}

\vspace{1cm}

\title{\Large \bf Inclusive Radiative $J/\psi$ Decays}



\author{D.~Besson}
\affiliation{University of Kansas, Lawrence, Kansas 66045, USA}
\author{T.~K.~Pedlar}
\affiliation{Luther College, Decorah, Iowa 52101, USA}
\author{D.~Cronin-Hennessy}
\author{K.~Y.~Gao}
\author{J.~Hietala}
\author{Y.~Kubota}
\author{T.~Klein}
\author{B.~W.~Lang}
\author{R.~Poling}
\author{A.~W.~Scott}
\author{P.~Zweber}
\affiliation{University of Minnesota, Minneapolis, Minnesota 55455, USA}
% ...
\collaboration{CLEO Collaboration}
\noaffiliation


\begin{abstract}
Using data taken with the CLEO-c detector at the Cornell Electron Storage
Ring, we have investigated the direct photon momentum 
spectrum in the decay
$J/\psi({\rm 1S}) \to \gamma gg$, via the ``tagged'' process:
$e^+e^-\to\psi$(2S); $\psi$(2S)$\to J/\psi~\pi^+\pi^-$;
$J/\psi\to\gamma$+X.
Including contributions from 
two-body radiative decay processes, we find 
the ratio of the inclusive direct photon
branching fraction to that of the dominant 
three-gluon branching fraction 
($R_\gamma=B(gg\gamma)/B(ggg)$) to
be \Rg,
where the errors shown are
statistical, systematic, and the model-dependent uncertainty 
related to the extrapolation to zero photon
energy.
The shape of the
scaled photon energy spectrum in 
$J/\psi\to gg\gamma$ is observed to be very similar
to that of
$\Upsilon\to gg\gamma$. The $R_\gamma$ value obtained is
roughly consistent with that expected by a simple quark-charge
scaling ($R_\gamma\sim(q_c/q_b)^2$)
of the value determined at the $\Upsilon$({\rm 1S}), but somewhat higher
than the value expected from the running of the strong coupling constant.
\end{abstract}
\maketitle

\section{Introduction}
According to the Okubo-Zweig-Iizuka (OZI) rule, 
the preferred decay mode for charmonium
would be through the production of a $D{\overline D}$ meson pair.  
For resonances
below the $\psi$(3770) 
however, this is not energetically possible.  Thus the decay
of the $J/\psi$(1S) meson must proceed through
OZI-suppressed channels.
The three lowest-order decay modes
of the $J/\psi$ meson are the three-gluon ($ggg$), 
virtual photon (vacuum
polarization) decays $J/\psi\to l^+l^-$
and $J/\psi\to q{\overline q}$ with
a branching fraction given in terms of 
${\cal R}_{QCD}\equiv (e^+e^-\to q{\overline q})/(e^+e^-\to\mu^+\mu^-)$ as
 $\sim {\cal R}_{QCD}\times{\cal B}(J/\psi\to l{\overline l})$,
and two gluon plus single photon ($gg\gamma$)
modes. 
For the $\psi$(2S)
and $\psi$(3770)
resonances, direct radiative transitions, both electromagnetic and
hadronic, as well as decays to open charm (for the $\psi$(3770)),
compete with these annihilation modes and therefore reduce
the $gg\gamma$ branching fraction.

\subsection{Inclusive Total Rate}
Since
$\Gamma_{ggg} \propto \alpha_s^3$ and $\Gamma_{gg\gamma} \propto  
\alpha_s^2\alpha_{em},$
the ratio of the branching fractions for the $ggg$ and $gg\gamma$ 
decay modes for heavy quarkonia (also equal to the ratio
of experimentally measured events $N_{gg\gamma}$ and 
$N_{ggg}$, respectively) is expected to follow \cite{r:Brod-Lep-Mack}:
\begin{equation}
R_{\gamma} \equiv \frac{B_{gg\gamma}}{B_{ggg}} = 
                  \frac{N_{gg\gamma}}{N_{ggg}}           =
	\frac{36}{5}q_c^2\frac{\alpha_{em}}{\alpha_s}[1+(2.2\pm0.6)\alpha_s/\pi]
\end{equation}
In this expression, the charm quark charge $q_c=2/3$.
Alternately, one can normalize to the well-measured dimuon channel
and cancel the electromagnetic vertex:
${B_{gg\gamma}/B_{\mu\mu}} \propto \alpha_s^2$.
In either case, one must define the momentum scale ($Q^2$) appropriate for this process.
Although the value $Q^2\sim M^2$ seems natural, the original prescription of Brodsky, Lepage and McKenzie (``BLM'' \cite{r:Brod-Lep-Mack}) gave $Q=0.157M_{\Upsilon({\rm 1S})}$ for (the less-relativistic) $\Upsilon$(1S)$\to gg\gamma$. 
Alternative prescriptions for the appropriate value of $Q^2$ have also been
suggested \cite{r:KMR}.

\subsection{Energy and angular spectrum shapes}
Calculations of the direct photon energy spectrum were originally
based on 
decays of orthopositronium 
into three photons, leading to the expectation that
the $J/\psi$ direct photon energy 
spectrum should rise linearly with $z_\gamma (\equiv 2E_\gamma/M_{J/\psi})$
 to the kinematic
maximum ($z_\gamma\to 1$); 
phase space considerations lead to a slight enhancement 
at $z_\gamma$=1 \cite{r:lowest-qcd}. 
The angular distribution for the decay of a 
polarized vector into three massless vectors is, in principle,
directly calculable. Thus,
for direct radiative decays $J/\psi\to gg\gamma$,
theory prescribes
the correlation of $z_\gamma$ with
photon polar angle $\cos\theta_\gamma$,
defined relative to the beam axis.
K\"oller and Walsh considered the angular
spectrum in detail \cite{r:kol-walsh}, 
demonstrating that, if the
parent is polarized along the beam axis, then, as the
energy of the
most energetic primary (photon or gluon)
in $J/\psi\to\gamma gg$ or $J/\psi\to ggg$ 
approaches the beam energy, the event axis tends
to increasingly align with the beam axis: 
$z_\gamma\to 1$ corresponds to $\alpha(z_\gamma)\to 1$ for
an angular distribution specified as
$dN/d\cos\theta_\gamma\sim 1+\alpha(z_\gamma)\cos^2\theta_\gamma$.
We note that, according to the
K\"oller-Walsh prescription, the value of
$\alpha(z_\gamma)$ for intermediate values, where most of the
events occur, is relatively small (0.2). Only for $z>0.9$ is the
forward peaking of the photon angular distribution noticeable.

Previous analyses of the direct photon spectrum
in heavy quarkonium decay selected a fiducial angular
region and integrated over $\cos\theta_\gamma$. In this analysis, we will take
advantage of the expected correlations between
$\cos\theta_\gamma$ and $z_\gamma$ to improve the
statistical precision of the extracted branching fraction.
There is nevertheless still some model dependence in the 
extrapolation down to $z_\gamma\to 0$.

\subsection{Previous Work}
Garcia and Soto (GS) \cite{r:GarciaSoto07} have
performed the most recent calculation of the 
expected direct photon spectrum in $J/\psi$
decays, using an approach similar to that
applied by the same authors for the case of 
$\Upsilon$(1S)$\to gg\gamma$ \cite{GSU1S}.
They model the
endpoint region by combining Non-Relativistic QCD
(NRQCD) with Soft Collinear-Effective Theory (SCET),
which facilitates calculation of the spectrum of the collinear gluons
which occur as $z_\gamma\to 1$. 
Both color-octet and color-singlet contributions
must be explicitly calculated and summed. The calculations are
very sensitive to the handling of the octet contribution, and
limit the momentum interval over which the theory is considered
`reliable' to $0.4\le z_\gamma\le 0.7$.
At low energies (defined
as $z_\gamma<$ 0.45 for the case of
the $\Upsilon$), the
so-called ``fragmentation'' photon component, due to photon
radiation from final-state quarks, dominates.

Although inclusive radiative decays have received
considerable experimental attention in the case of 
$b{\overline b}$
\cite{shawn,r:Albrecht-92,r:Csorna,r:Albrecht-87,r:Bizeti,r:nedpaper},
the 
$c{\overline c}$ system has had only one prior measurement, 
by the MARK-II Collaboration in 1981 \cite{r:MARKII-81}.
The MARK-II analysis, which utilized a calorimeter with resolution
$\sigma_E/E\sim 0.12/\sqrt{E({\mathrm GeV})}$  
resulted in a measurement for the inclusive
partial
branching fraction 
${\cal B}(J/\psi\to\gamma X$)=$(4.1\pm0.8)$\%
limited to the range $z_\gamma>$0.6. 
Although the authors do not
explicitly quote a value for $R_\gamma$ in their original reference,
we can estimate an implied $R_\gamma$ value
assuming that $z_\gamma>0.6$ constitues 45\% of the total spectrum
(over the full $\cos\theta_\gamma$ range, based on
results obtained for bottomonium), and
using ${\cal R}_{QCD}^{ECM=3.1~{\mathrm{GeV}}}$=2.1 and
${\cal B}_{\mu\mu}$=0.059, such that ${\cal B}(ggg)$
+ ${\cal B}(gg\gamma)$=0.69, yielding
$R_\gamma={\cal B}(gg\gamma)/{\cal B}(ggg)$
$\sim (14.6\pm2.8)\%$. 
The MARK-II direct photon spectrum peaked at $z_\gamma\sim0.6$,
inconsistent with expectations based on orthopositronium
decay, but consistent with later bottomonium spectra.

\section{Detector and Event Selection}
The CLEO-c detector is essentially identical to the previous
CLEO~III detector, with the exception of a modified
innermost charged particle tracking system.
Elements of the detector, as well as performance characteristics,
are described in detail elsewhere  \cite{r:CLEO-II,r:CLEOIIIa,r:CLEOIIIb}. 
Over the kinematic regime of
interest to this analysis, the
electromagnetic shower energy resolution is approximately 2\%.
The tracking system, 
RICH particle identification system, 
and electromagnetic
calorimeter are all contained within a 1 Tesla superconducting
solenoid.

In the absence of dedicated $J/\psi$ data collected
at CLEO-c, we use the cascade
decay chain $\psi(2S)\to J/\psi\pi^+\pi^-$; 
$J/\psi\to gg\gamma$. Our data sample corresponds to 
approximately 27$\times 10^6$ $\psi$(2S) decays\cite{Npsiprime} 
collected with the
CLEO-c detector (our ``primary'' 
data sample, divided into two sub-samples of data taken
approximately three years apart); a much smaller (``secondary'') 
sample of slightly more
than $10^6$ $\psi$(2S) decays collected with the CLEO-III
detector is used for cross-checks.
To ensure maximal efficiency,
minimal event selection requirements are imposed on our
candidate direct photon sample -- we require only that
candidate
events have at least two high-quality 
charged tracks (the two transition 
charged pion candidates)
and no identified
lepton charged tracks. Our lepton veto is effective in 
suppressing contamination from $J/\psi\to l^+l^-$.
Monte Carlo simulations
indicate that the trigger efficiency for events having
two transition charged pions and a 
fiducially-contained direct photon
with $z_\gamma>$0.3 is $>$99\%. 

\section{Analysis}

\begin{figure}[htpb]
\centerline{\includegraphics[width=9cm]{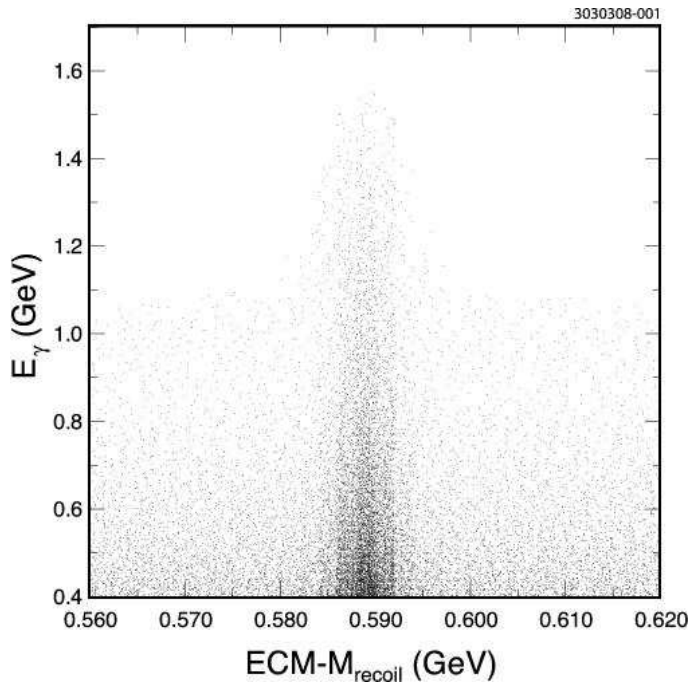}}
\caption{Center-of-mass energy 
minus dipion recoil mass (horizontal) vs. photon energy
(vertical), $\psi$(2S) data.}
\label{fig:mxx-Egamma}
\end{figure}

To obtain $R_{\gamma}$, we must determine separately 
the number of direct photon
events and the number of three gluon events. 
By using the 
decay chain $\psi(2S)\to J/\psi\pi^+\pi^-$; 
$J/\psi\to gg\gamma$, we circumvent initial state radiation backgrounds in
our photon sample. 
The dominant background for $z_\gamma<$0.6
is primarily from $\pi^0\to\gamma\gamma$ and $\eta\to\gamma\gamma$.
Photon selection requirements are essentially the same
as applied for our study of $\Upsilon\to gg\gamma$ \cite{shawn}. Namely,
we require showers detected in the electromagnetic
calorimeter with energy deposition characteristics
consistent with those expected for true photons,
and which are well-isolated from both charged tracks as well
as other showers. 
Since the $\pi^0$ signal to noise is high, we will, in contrast
to our previous analysis,
require
that a candidate high-energy photon not combine
with another high-quality photon to give an
invariant mass within $\pm$6 MeV
from the nominal $\pi^0$ mass $m_\pi^0$. However, we
do not impose an $\eta$ veto given the worse signal to noise and
the greater likelihood of incorrectly vetoing a true
direct photon. 
The scatter plot of the
raw candidate photon energy
vs. ``shifted''
dipion recoil mass ($ECM-M_{\mathrm{recoil}}$)
is shown in Figure \ref{fig:mxx-Egamma}.

This spectrum includes contributions from each of the three main decay
modes of the $J/\psi$: 
three-gluon, vacuum polarization, and direct photon. Photons
from nonresonant processes below the resonance, 
$e^+e^- \rightarrow q\bar{q}(\gamma)$, 
also contribute to
this spectrum. However, since the charmonium peak cross-section is a factor
of 50 times larger than the continuum cross-section, and 
we perform a dipion-sideband subtraction 
to obtain the ``tagged''
direct photon spectrum from the $J/\psi$, we will, in what follows, ignore
this continuum contribution. For the sideband subtraction,
the signal region is defined as a dipion recoil region within
$\pm$10 MeV/$c^2$ of the canonical $J/\psi$ mass; sidebands
are defined as the regions 10--40 MeV/$c^2$ from the canonical
$J/\psi$ mass. 

Knowing the dipion four-vector allows us
to work in the rest frame of the $J/\psi$ itself; in what follows,
unless otherwise indicated, the energy spectra presented
correspond to this case.

\section{Background Photon Subtraction And Signal Estimate}
In the previous bottomonium analysis, two parameterizations of the
background were used -- one was based on the ``pseudo-photon''
technique which has been used in three previous CLEO 
analyses \cite{r:Csorna,r:nedpaper,shawn},
as well as the original MARK-II analysis,
and the other
used a simple exponential parametrization of the background
under the direct photon signal. The latter suffers from integrating
over the correlations between $\cos\theta_\gamma$ and
$z_\gamma$ and is therefore not used in our current analysis.
To model the production of $\pi^0$ and
$\eta$ daughter photons, 
redundant estimators were
employed in this analysis, which we detail below. 
Unlike the bottomonium analysis,
we do not simulate $\omega \rightarrow \pi^0 \gamma$, and
$\eta' \rightarrow \gamma (\rho, \omega, \gamma)$ contributions.
JETSET \cite{r:LUND} indicates that these should be
smaller than for the 
$\Upsilon$. Numerically, the fraction of 
all $z_\gamma>$0.4 photons having
$\omega(\eta'$) parentage is 2.2\%(0.8\%) according
to the LUND event generator, somewhat 
below our typical systematic errors.
The non-photonic contribution to our final 
candidate shower sample, due almost exclusively
to $K_L$ and ${\overline n}$ interactions in the
calorimeter, is estimated
from Monte Carlo simulations to be 1.5\%.
We note that, since the background shape is fixed
during signal extraction, but the
normalization allowed to float, such
non-direct photon contamination is
largely absorbed into the eventual
background estimate.

\subsection{``Pseudo-photons''}
As a first estimate of the non-direct photon background,
we took advantage of the 
expected similar kinematic distributions between charged and
neutral pions, as dictated by isospin symmetry. 
Although isospin
will break down both when there are decay processes which are not
isospin-symmetric in their final states
($J/\psi\to\gamma\eta$, $\eta\to\pi^0\pi^0\pi^0$) or
when the available fragmentation phase space is
comparable to $M_\pi$, in the
intermediate-energy regime 
we expect isospin to be reliable. Over the kinematic
regime relevant for this analysis,
Monte Carlo studies indicate consistency (to within
$\sim$5\%) with the
naive expectation that
there should be half as many neutral pions as charged pions,
with similar momentum-dependent angular distributions.
However, unlike the case for
the $\Upsilon$, the region $z_\gamma\to$1 has large
contributions from two body radiative decays. Specifically,
the $\pi^0:\pi^\pm$ ratio grows in this regime due to decays
such as
$J/\psi\to\gamma\eta$ ($\eta\to 3\pi^0$) and
$J/\psi\to\gamma\eta'$.

A pseudo-photon spectrum ($d^2N/(dz_\gamma d\cos\theta_\gamma$)) is
constructed using charged tracks with particle identification
information consistent with pions to model the 
spectra expected
for $\pi^0$'s and $\eta$'s. In both cases, we use the
Monte Carlo prescribed $\pi^0:\pi^\pm$ or
$\eta:\pi^\pm$ ratios,
taking into account the variation in these ratios with momentum.
Momentum-dependent
corrections are also applied to account for 
non-pion charged kaon, proton, and lepton
fakes in our sample, as well as the finite charged
track-finding efficiency. Each of these
last two corrections are
of order 5\% and tend to offset each other.
We invoke isospin in making the assumption that the momentum-dependent
angular distributions of neutral pions follows that of charged pions.
Our Monte Carlo generator indicates that the momentum-dependent angular
distribution for $\eta$'s is also similar to charged tracks. 
We simulate the two-body decays $\pi^0\to\gamma\gamma$ and
$\eta\to\gamma\gamma$ in the rest frame of the candidate 
$\pi^0$ or $\eta$ parent
and boost the daughter pseudo-photons
into the lab frame 
according to the ($\pi^0$ or $\eta$)
momentum. 
The direct-photon finding efficiency $\epsilon_\gamma(E_\gamma)$
is applied to each daughter pseudo-photon to
determine the likelihood that that
photon will
populate our candidate direct photon spectrum. 
From Monte Carlo simulations, $\epsilon_\gamma\sim0.85$
over the
kinematic and geometric fiducial interval defined in this
measurement.
Finally,
``found'' pseudo-photons are smeared in energy and angle
by the known resolutions. 

\begin{figure}[htpb]\centerline{\includegraphics[width=9cm]{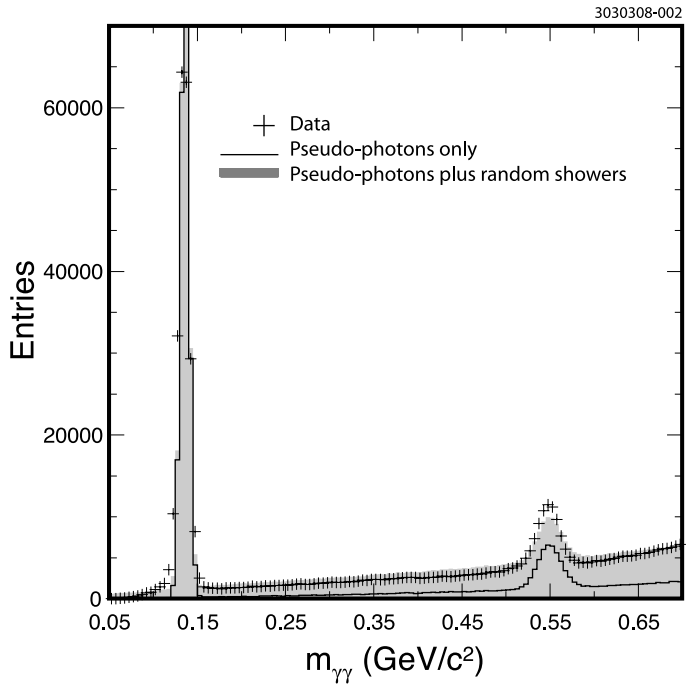}}\caption{Two-photon invariant mass combinations in data (crosses) vs. pseudo-photon simulation. Shaded histogram indicates pseudo-photon simulation after adding combinations of simulated $\pi^0$ and $\eta$ daughters with ``excess'' showers in data. Normalization is absolute.}\label{fig:pi0_check}\end{figure}

To check our procedure, 
we have compared the data $\gamma\gamma$ invariant mass plot
with the
pseudo-photon $\gamma\gamma$ invariant mass plot.
To enhance statistics, we use all photons with $E_\gamma>$0.2
GeV ($z_\gamma>0.13$). At such relatively low photon
energies, the number of accepted showers not having $\pi^0$ or
$\eta$ parentage is considerable, so we must also
add to our $M_{\gamma\gamma}$ spectrum combinations of
``found'' $\pi^0$ or $\eta$ daughter pseudo-photons with
``excess photons'', using a Monte Carlo-based momentum-dependent
factor.
We obtain a level of agreement (better than 10\%) consistent
with our previous $\Upsilon$(1S) analysis \cite{shawn}.
The comparison between our pseudo-photons and data is shown in Fig. \ref{fig:pi0_check}.

\subsection{Background Estimate from Monte Carlo Simulations}
Second, we use the Monte Carlo simulation of generic
$J/\psi$ decays, based on the JETSET 7.4 event generator,
to provide an estimate of the
background to the non-direct photon signal, including all sources.
This estimate implicitly includes all the 
corrections (photon efficiency, tracking efficiency and 
fake rates, $\pi^0:\pi^\pm$ ratio, hadronic showers, etc.) which must
be explicitly evaluated in the previous approach. 
No additional corrections are therefore applied in this case.

\subsection{Background using $\chi_{c0}$ decays}
\begin{figure}[htpb]
\centerline{\includegraphics[width=9cm]{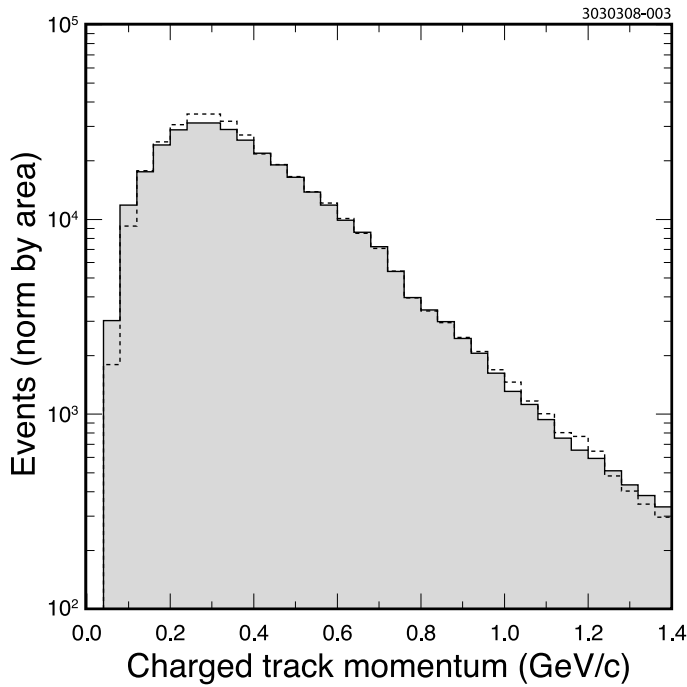}}
\caption{Comparison of $\chi_{c0}\to\pi^\pm$ (shaded) vs.
$J/\psi\to\pi^\pm$ (dashed) inclusive charged track spectra.}
\label{dndp}
\end{figure}

A third estimate of the background is obtained from decays of the
$\chi_{c0}$(3415), which is produced via the transition:
$\psi$(2S)$\to\chi_{c0}\gamma$ with the emission of a photon
of energy 270 MeV. Since the $\chi_{c0}$ is relatively
wide ($\Gamma\sim$10 MeV), it has a relatively small 
radiative decay branching fraction to the $J/\psi$ (${\cal B}$=$1.32\pm0.11$\%)
and therefore dominantly decays via two-gluon 
intermediate states. To the
extent that two-gluon fragmentation is similar to three-gluon
fragmentation \cite{chib-study}, we can therefore use the data photon background
produced in association with an observed $\psi$(2S)$\to\chi_{c0}\gamma$
transition photon candidate to estimate the non-direct photon
background to the $\psi\to gg\gamma$ photon energy spectrum.
Comparison of the charged track spectra 
for sideband-subtracted
$\chi_{c0}$ vs. $J/\psi$ decays indicates that the kinematics
of the former two-gluon decays are similar to the latter three-gluon
decays in this case (Figure \ref{dndp}).

To further suppress any possible $\chi_{c0}\to J/\psi\gamma$
cascade contamination, we 
veto events which contain a high-quality
photon candidate of energy $322\pm$20 MeV. 
A signal
region is defined around the $240-290$ MeV transition photon
energy range and the photon energy spectrum in coincidence
with the candidate transition photon (after sideband subtraction, with
sidebands taken an additional 25 MeV on each side of the signal
region)
is then used as our last
background-estimator for our final fits.

\subsection{Polarization of the parent $J/\psi$}
In principle, the dipion transition can be either S- or D-wave, as allowed
by parity conservation. The BES Collaboration have studied 
the angular distributions for this process \cite{BESdipion} 
and find a best-fit value for the 
D-wave to S-wave amplitude of 0.18$\pm$0.04. 
If the decay is all D-wave, then the $1+a\cos^2\theta_\gamma$ distribution
expected for two-body
$\gamma$+pseudoscalar decays softens to $a\sim0.07$ compared to
$a=1$ for S-wave. (We use the symbol $a$ to designate the angular
distribution of the radiative daughter photon specifically in two-body
radiative decays.) 
Similarly, there is some uncertainty in the
angular distribution of the direct photon signal itself, characterized
by the inclusive spectral parameter $\alpha(z_\gamma)$.
To accommodate this, we have done fits varying values of both
the two-body angular coefficient $\alpha(z_\gamma)$ and the
inclusive direct photon angular coefficient
$a$, and include the difference among them as a systematic
error.

\subsection{Fits and Signal Extraction}
After imposing our event selection and photon selection criteria,
we are left with the two-dimensional 
candidate direct photon scaled
energy vs. polar angle
distribution.
We perform 
two-dimensional fits comprised of the following components: a)
the background, which is modeled either using the pseudo-photon,
Monte Carlo-based, or $\chi_{c0}$-based backgrounds described above,
b) three two-body components of the direct photon signal: 
$\gamma\eta$ and $\gamma\eta'$ and
a wide resonance which corresponds to $\gamma\eta(1440)$, with shapes
determined from Monte Carlo simulation, and c)
a smooth signal component which has a shape in photon
energy taken from our previous
Upsilon decay measurement \cite{shawn}, and an angular 
distribution based on the K\"oller-Walsh prescription \cite{r:kol-walsh}. Ideally, 
we could 
avoid having to include a signal component. In such a case,
the
background subtraction
would directly determine the true underlying signal. However,
this can only be done if the background can be 
absolutely normalized with very 
high precision (much better than our $\sim$10\% 
overall background normalization
error). Unfortunately,
the statistics of the fit are largely driven by the low-$z_\gamma$ region, 
where the systematic uncertainty on the
$\pi^0$ background is largest.
Without inclusion of a signal component, the background normalization
would increase to saturate the low-$z_\gamma$ energy regime.

\begin{figure}[htpb]
\centerline{\includegraphics[width=9cm]{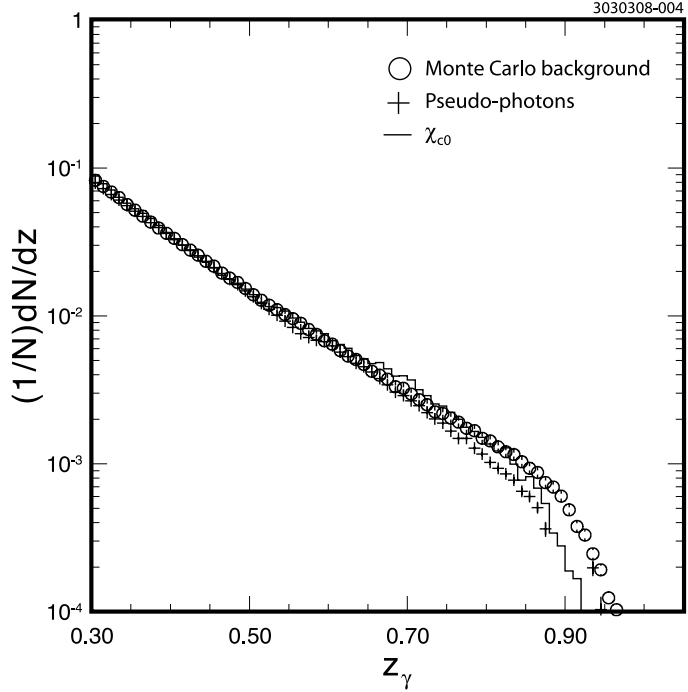}}
\caption{Comparison of background photon ($(1/N)dN/dz_\gamma$) 
spectra for three subtraction schemes
considered. 
Plus signs: pseudo-photon background; histogram: $\chi_{c0}$
background; circles: Monte Carlo simulations of background.}
\label{fig:projsgbkover-44-0.8-0.35}
\end{figure}

A comparison of the one-dimensional projections of the background 
$dN/dz_\gamma$ spectra
is shown in Figure \ref{fig:projsgbkover-44-0.8-0.35}.

\subsection{Validation of fit procedure using Monte Carlo simulations}
For the Monte Carlo simulations, 
we can fit the signal plot with a combination of the
tagged Monte Carlo simulation
signal plus either the
pseudo-photon or the $\chi_{c0}$-based background, then check the
signal normalization against the known 
number of signal direct photons in the Monte Carlo simulations. 
The $\chi_{c0}$ background estimator
results in 
a signal yield only 3\% larger than the known number of
Monte Carlo direct photons in the plot, while the 
pseudo-photon background estimator underestimates
the signal yield by $\sim$15\%.
We note that the agreement between
the signal yields obtained
from these two backgrounds in data is typically within 3\%.
As discussed later, we nevertheless add an additional systematic error (6\%) to
reflect this discrepancy observed in simulation.

\section{Results}
\begin{figure}[htpb]
\centerline{\includegraphics[width=12cm]{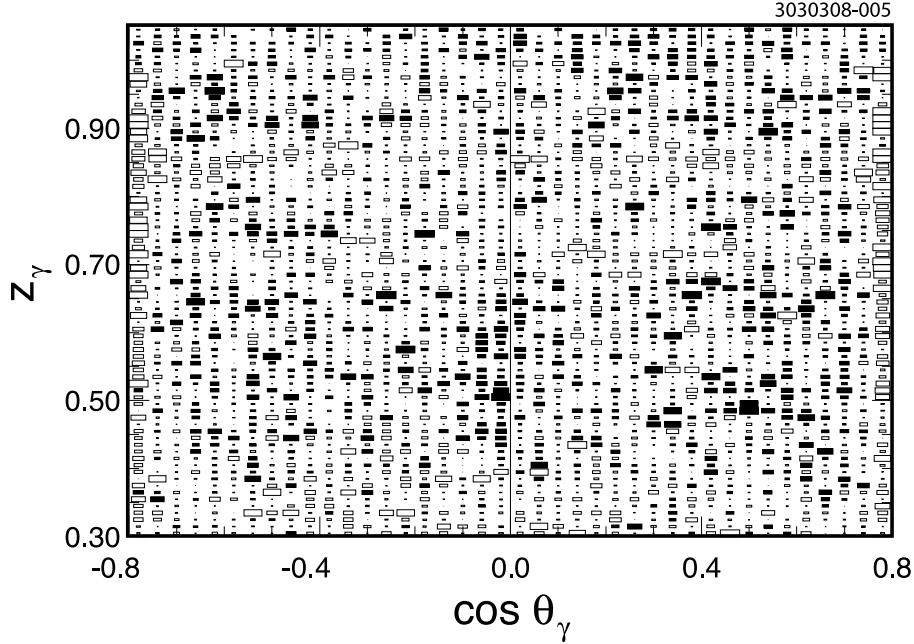}
}
\caption{Fit over $z_\gamma>0.3$, $|\cos\theta_\gamma|<$0.8 kinematic region
using pseudo-photon 
background. Shown are
normalized residuals, in units of statistical error per bin.
Black: data$>$fit; White: data$<$fit. Excess at large values of
$\cos\theta_\gamma$ attributed to QED processes producing 
charged leptonic tracks
at large dip angles, which are
(incorrectly) used as input to pseudo-photon 
generator. This results in excess pseudo-photons at high
values of $\cos\theta_\gamma$.}
\label{fig:data42a}
\end{figure}

\begin{figure}[htpb]
\centerline{\includegraphics[width=12cm]{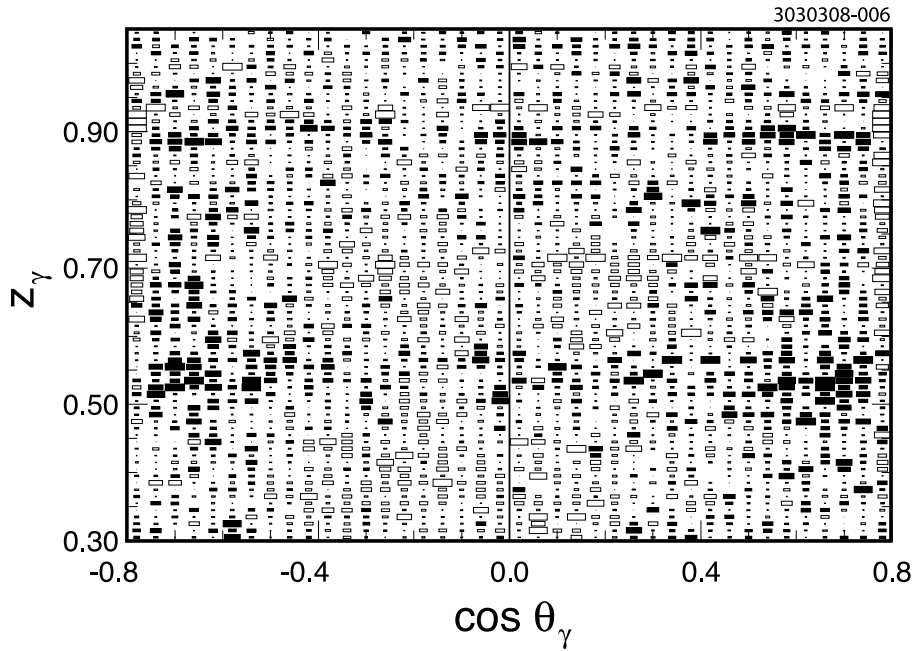}
}
\caption{Fit over $z_\gamma>0.3$, $|\cos\theta_\gamma|<$0.8
kinematic region, MC background. 
Black: data$>$fit; White: data$<$fit. Note apparent presence of
extra two-body component in data, at $z_\gamma\approx 0.5-0.55$.}
\label{fig:data42b}
\end{figure}

\begin{figure}[htpb]
\centerline{\includegraphics[width=12cm]{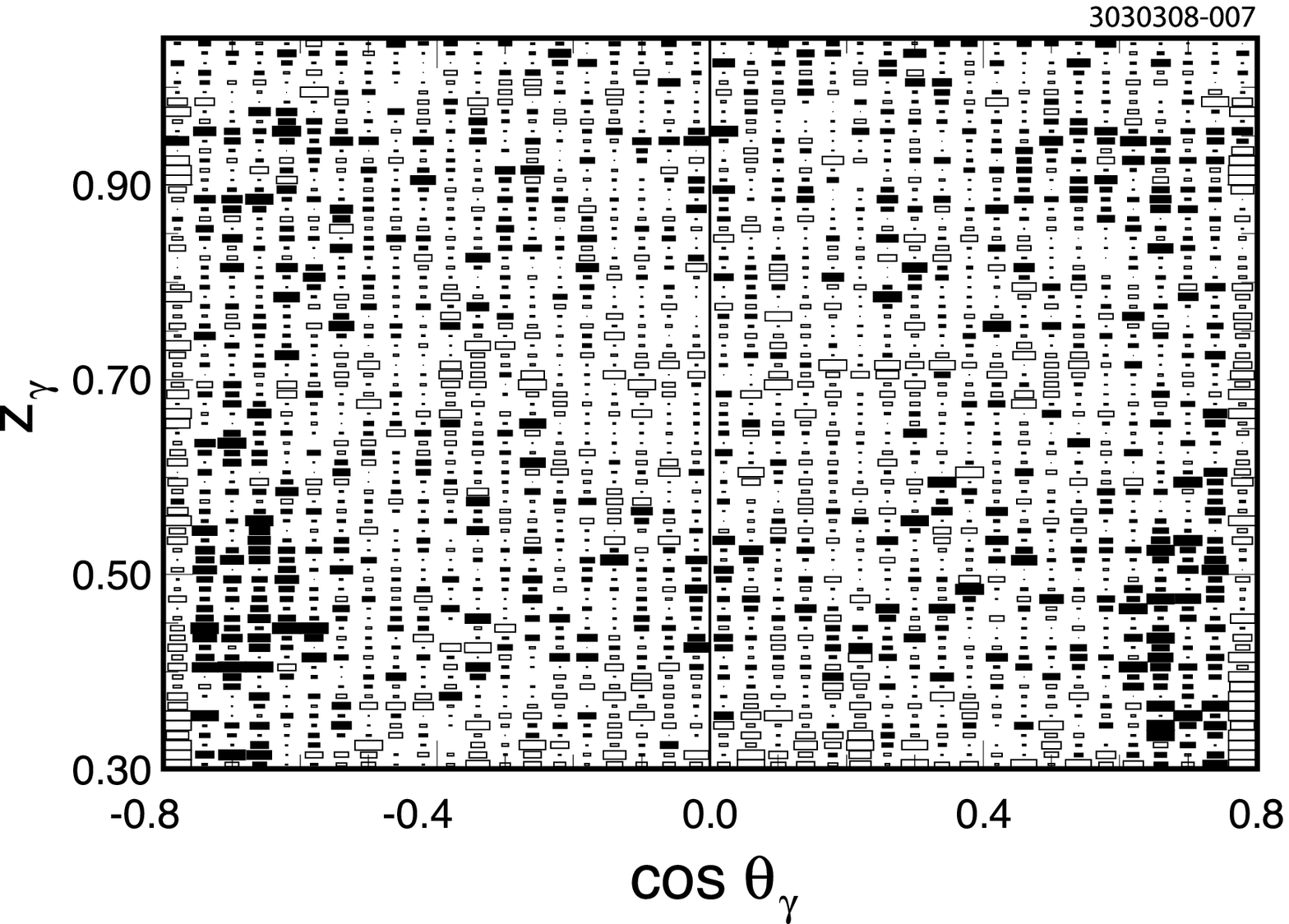}}
\caption{Fit over
$z_\gamma>0.3$, $|\cos\theta_\gamma|<$0.8 kinematic
region, $\chi_{c0}$ background. Black: data$>$fit; White: data$<$fit.
We again note apparent excess of data over background in high 
$\cos\theta_\gamma$ region around $0.4<z_\gamma<0.5$.}
\label{fig:data42c}
\end{figure}

Figs. \ref{fig:data42a}, \ref{fig:data42b} and 
\ref{fig:data42c} show sample fits over the
kinematic region $z_\gamma>$0.3 and
$|\cos\theta_\gamma|<$0.8, based on the
pseudo-photon model of the background, the Monte Carlo-based
model of the background, and the $\chi_{c0}$-based
model of the background, respectively.

\begin{figure}[htpb]
\centerline{\includegraphics[width=16cm]{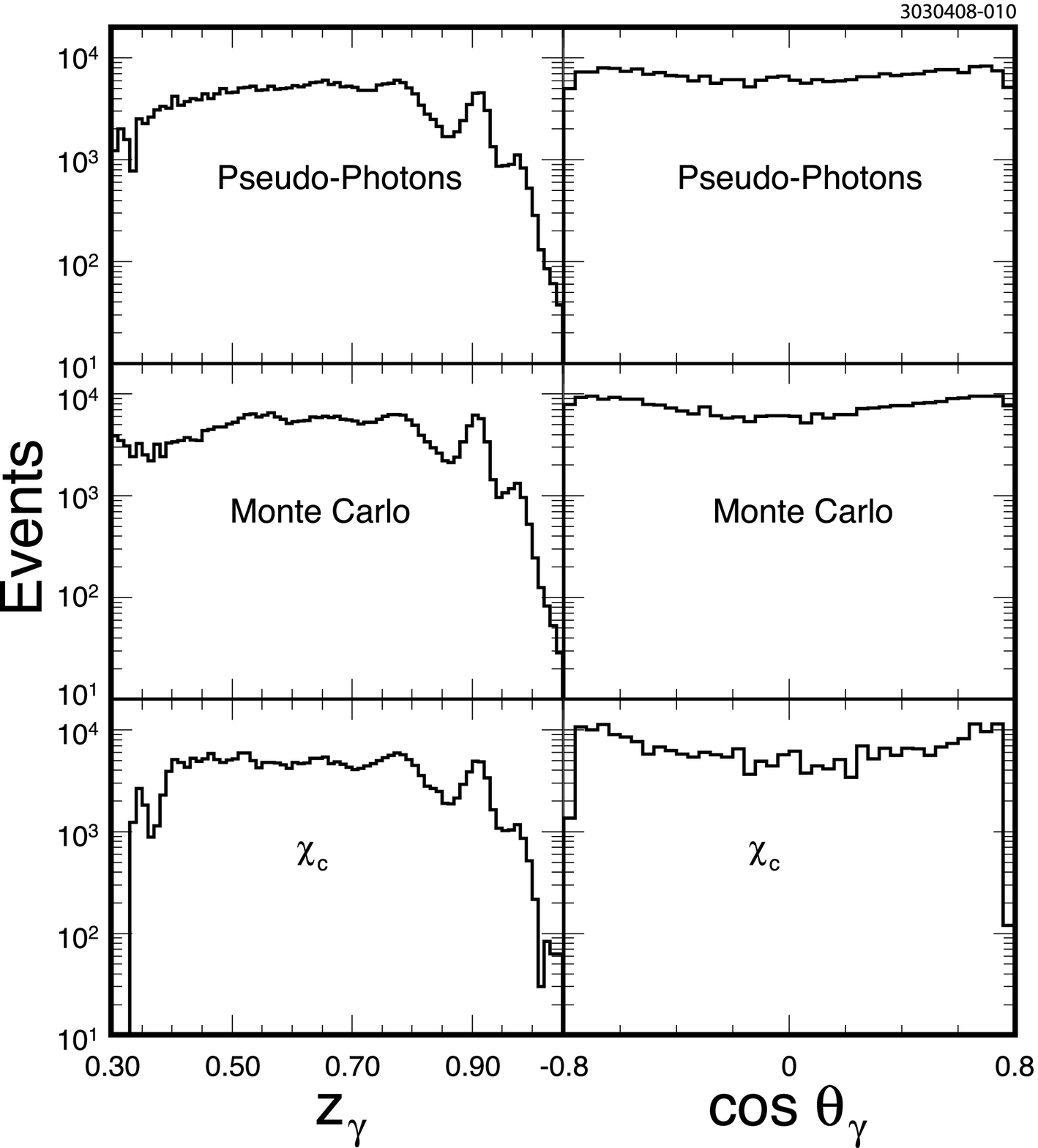}}
\caption{Background-subtracted
signal yield projections onto $z_\gamma$ and $\cos\theta_\gamma$ axes, 
for the three photon background estimators used in this analysis.}
\label{fig:SixPanel}
\end{figure}

\begin{figure}[htpb]
\centerline{\includegraphics[width=11cm]{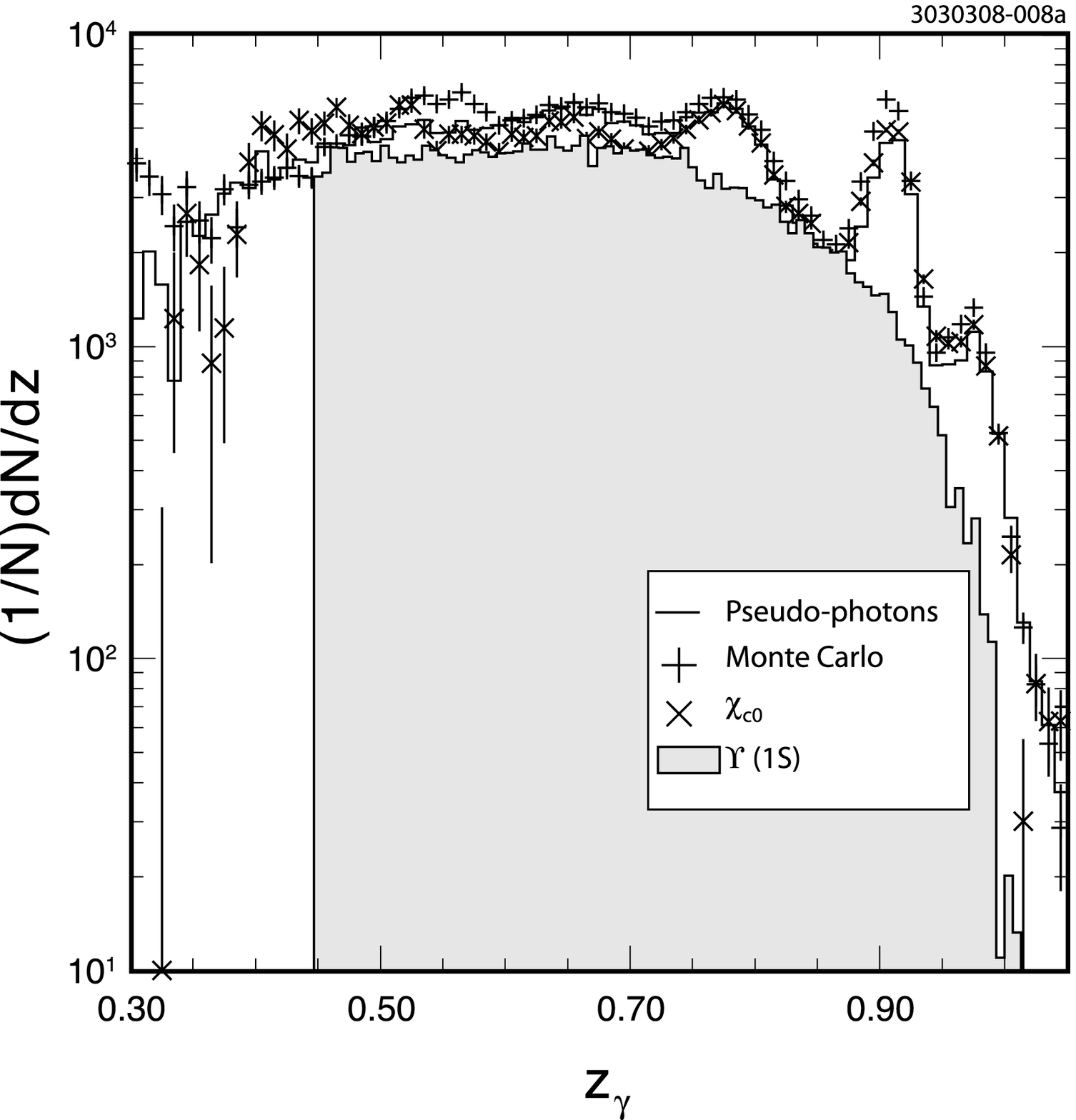}}
\caption{$z_\gamma>0.35$, $|\cos\theta_\gamma|<$0.8, 
background-subtracted direct photon 
energy spectra, using three different background subtraction schemes. Also overlaid (shaded histogram) is the experimental spectrum for the $\Upsilon$ spectrum.} 
\label{fig:moneyplot}
\end{figure}

Positive residuals (data in one bin
exceeds sum of fit contributions) are shown in black;
negative residuals (data in one bin is smaller than sum of fit 
contributions) are shown in white. Individual projections, onto the
scaled photon energy and photon polar angle axes, for the three background
models
separately, are presented in Figure \ref{fig:SixPanel}.
An overlay
of the background-subtracted spectra, for the three
background models employed, is shown in Figure \ref{fig:moneyplot} for 
our primary dataset. We observe reasonable agreement between the
three spectra over most of the kinematic regime considered.

Excluding electromagnetic transitions to other charmonium states,
two-body radiative
exclusive modes should be included in our
total $gg\gamma$ yield. 
Decays into
narrow $\eta$ mesons have large enough branching fractions 
(${\cal B}(\gamma\eta)=(9.8\pm1.0)\times 10^{-4}$ and
${\cal B}(\gamma\eta')=(4.71\pm0.27)\times 10^{-3}$)
so that they are clearly visible in the 
one-dimensional projection of the signal
spectrum. Rather than fixing these contributions, we have allowed
them to float and use the fitted area, corrected by efficiency, as a check
of the overall procedure. We obtain estimates 
for the two-body branching fractions of 
$(8.1\pm0.6)\times 10^{-4}$ and
$(4.98\pm0.08)\times 10^{-3}$ for our primary data 
sample
and 
$(8.8\pm2.1)\times 10^{-4}$ and
$(5.4\pm0.4)\times 10^{-3}$ for our lower statistics,
secondary data sample, where the errors presented are
statistical only. Systematic errors for this
coarse cross-check are likely to be
at least as large.

\section{Extraction of $R_\gamma$}
To determine $R_\gamma$, we calculate the ratio of true
$N_{gg\gamma}$ and $N_{ggg}$ events (Eq. (1)). The former quantity is
the number of observed direct photon candidates, corrected by
the photon-finding efficiency and the kinematic acceptance.
The latter quantity is the total number of 
$\psi$(2S)$\to J/\psi\pi^+\pi^-$ events estimated from our
inclusive dipion recoil mass spectrum, minus contributions from
$J/\psi\to gg\gamma$ (implicitly including all two-body radiative components), 
$J/\psi\to l^+l^-$, and
$J/\psi\to q{\overline q}$. 
The two-track trigger efficiency
is $>$99\% for all these processes. The hadronic event
reconstruction efficiency is also $>$99\%; due to our
lepton veto, the acceptance for dileptonic decays is less
than 5\%.

Table \ref{tab:toggle} shows the range of values obtained for 
two sub-samples of the primary data, taken approximately
three years apart 
(presented as ``sub-sample-1/sub-sample-2'')
 with varying definitions of the
signal region. 
\begin{table}[htpb]
\caption{Summary of values of $R_\gamma$ using varying
kinematic regions for fits, different datasets and 
background subtraction schemes. First number is 
primary sub-sample 1, second is primary sub-sample 2.
Numbers presented are simple averages of the values
obtained using the three different background estimators. 
The number in parenthesis following each
average corresponds to the rms of the difference among the
three background estimators and is therefore indicative
of the (dominant) signal estimation systematic error.
By comparison, statistical errors are typically
of order 1\% for sub-sample 1 and 0.2\% for sub-sample 2.
Our final quoted $R_\gamma$
result is a weighted average of the presented values.}
\begin{tabular}{ccc} \hline\hline
 & $|\cos\theta_\gamma|<$0.7 & $|\cos\theta_\gamma|<$0.8 \\ \hline 
$z_\gamma>0.30$      & 0.1367(88)/0.1398(78) & 0.1338(97)/0.1371(85) \\
$z_\gamma>0.45$   & 0.1320(110)/0.1362(94) & 0.1332(89)/0.1403(98) \\ \hline
\end{tabular}
\label{tab:toggle}
\end{table}

\section{Systematic errors}
\begin{figure}[htpb]\centerline{\includegraphics[width=9cm]{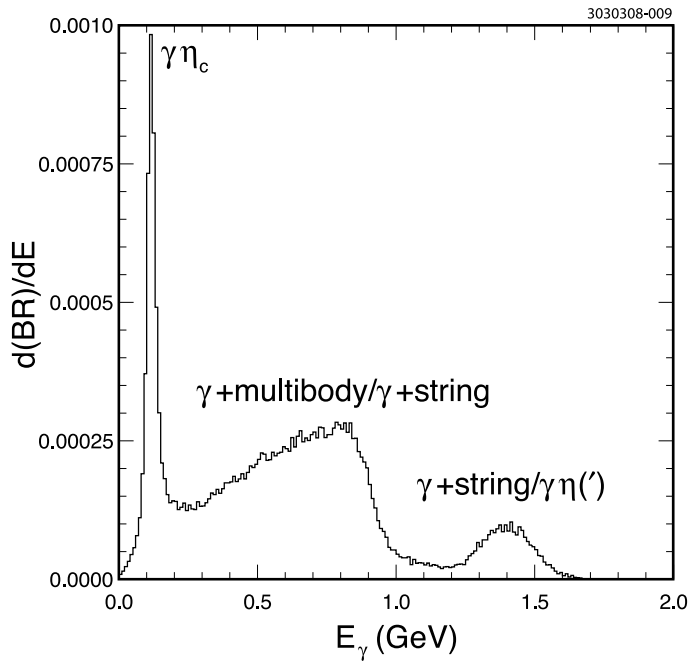}}
\caption{Monte Carlo simulations of direct photon $J/\psi$ radiative decay photon spectrum.}\label{fig:psi2gamX_glevel_MC}\end{figure}

We identify and estimate systematic errors in our $R_\gamma$
determination as follows:
\begin{enumerate}
\item The uncertainty in the S:D 
admixture of the dipion transition,
in principle, affects the 
angular distribution ($\sim 1+a\cos^2\theta_\gamma$) 
and therefore the
acceptance for the two-body radiative component in our fits,
as well as the angular distribution, and acceptance for the primary direct
photon signal component ($\sim 1+\alpha(z_\gamma)\cos^2\theta_\gamma$). 
Although both S-wave and the dominant allowed D-wave
transition amplitudes leave
the daughter $J/\psi$ polarized along the beam axis, we nevertheless
allow for possible contributions due to D-wave amplitudes resulting
in the daughter $J/\psi$ polarized transverse to the beam axis.
Varying $a$ between 0.7 and 1.0 for the two-body modes
results in a 1\% change in the extracted direct
photon yield; varying $\alpha(z_\gamma)$ between the value
prescribed by K\"oller-Walsh and $\alpha$=0 for
all photon momenta results in a 3\% lower value for the extracted
direct photon yield. We attribute this to the larger saturation of
the signal region by background which results when the signal
photon angular distribution is taken to be flat in angle.
We assume a systematic uncertainty of
3\% due to this source.
\item The uncertainty in the contribution to the signal 
due to non-photon showers, based on Monte Carlo modeling of 
signal and background decays,
is estimated to be 1.5\%. 
\item Uncertainty in the number of 3-gluon events is obtained by
subtracting from the total number of dipion tags the number of
$J/\psi\to\gamma^*\to q{\overline q}$ events, the number of
signal $J/\psi\to\gamma^*\to gg\gamma$ events, and the number of
$J/\psi\to\gamma^*\to l^+l^-$ dileptonic decay events which pass our cuts.
The statistical error on the branching fraction for
$J/\psi\to\gamma^*\to q{\overline q}$ is very small 
(${\cal B}=13.50\pm0.30$\%) \cite{PDG07}, as is the
error on the dileptonic branching fraction ($5.94\pm0.06$\%).
The fraction of dileptonic events which pass
our cuts is also small ($\le$5\%), as is the statistical
error on the number of $gg\gamma$ events in our sample. 
The total systematic error on our calculated $ggg$ yield
due to the non-$ggg$ subtractions is largely due to
the uncertainty in $R_\gamma$ and determined to be $\sim$2\%.
\item The trigger efficiency systematic
error in the ratio is $\le$1\%.
\item Background normalization and background 
shape uncertainty are evaluated by examining
the agreement between the direct photon yield obtained
using the three different background estimators.
We point out that these
three techniques sample very different methods of background-estimation.
The $\chi_{c0}$ subtraction background estimate, e.g., is 
insensitive to the uncertainty in the overall 
photon-finding efficiency. Given the observed agreement
across momentum, we infer that the pseudo-photon technique is
least likely to be sensitive to $\pi^0$/$\eta$
modeling uncertainties. 

However, we observe that the fits follow a generally
consistent pattern. Although the data-driven fits (pseudo-photon and 
using the $\chi_{c0}$ background) are generally 
consistent with each other at the 3\% level, the average of the
data-driven fits is consistently lower (by $\sim$15\%)
than the Monte Carlo-background subtracted spectra.
The overall rms of the signals obtained using the three background-estimators
is 7\% (see Table \ref{tab:toggle})
and we assign this value to the corresponding systematic error.
\item The uncertaingy in the absolute photon-finding efficiency is estimated at 2\%.
\item Sensitivity to the selection of signal and sideband
regions in the dipion mass spectrum is estimated by increasing the
nominal `signal' recoil mass interval by 25\% and decreasing the nominal
`sideband' recoil mass interval by 25\%, indicating a systematic error
$<$2\%.
\item The difference in our calculated
value of $R_\gamma$ between imposing vs. not imposing the
$\pi^0$ veto is found to be about 3\%.
\item For the pseudo-photon subtraction only, the
sensitivity to the assumed 
$\pi^0:\pi^\pm$ ratio was estimated by comparing the results based on 
the Monte Carlo-prescribed ratio vs. 
a constant value of 0.5. This results in a variation of 3\%
in $R_\gamma$.
\item Possible continuum QED contamination should be subtracted
out via the dipion sideband subtraction, although we do
rely on Monte Carlo simulations to quantify the background from
processes such as $\psi$(2S)$\to J/\psi\pi^+\pi^-$; $J/\psi\to l^+l^-\gamma$. 
Our results
with very strict QED suppression vs. no QED suppression vary
by 1\%. We conservatively assign a 1\% systematic error due
to our uncertainty in this background.
\item As described
previously, we have compared the signal yield with the ``true''
signal yield using a Monte Carlo-only study, in which the number
of simulated signal photons are known. Unfortunately, the JETSET 7.4 Monte 
Carlo simulated spectrum is entirely two-body and quasi-two-body,
and does not reproduce data well (Figure \ref{fig:psi2gamX_glevel_MC}). 
As outlined previously, we find
that our
average extracted signal yield 
is smaller than the true signal magnitude in
Monte Carlo simulations by 6\%, and conservatively
(since this
error likely is somewhat redundant with the systematic error assessed
by the spread in $R_\gamma$ values obtained using the three different
subtraction schemes)
include this as an additional systematic error. 
\end{enumerate}

A separate, additional 
systematic error must be included to account for the uncertainty
in the extrapolation to $z_\gamma$=0.
For this, we compare the values obtained 
assuming a linear
extrapolation from $z_\gamma$=0.3 to $z_\gamma$=0 vs. an
extrapolation based on the shape of the spectrum observed in
the case of the $\Upsilon$.
The difference between the yields for these two extrapolations
is $\le$3\%.

Table \ref{tab:worldresults} 
compares the results of this analysis with those obtained by previous
experiments.
Although the statistics are poorer, the
older CLEO-III data (our cross-check sample) 
gives results which are consistent with the CLEO-c
results ($0.132\pm0.008\pm0.013$, where the first error is
statistical and the second represents the spread in the measured
values obtained using the three different 
background subtraction schemes).

\begin{table}
\begin{center}
\begin{tabular}{cc}
\hline \hline
 Experiment                 & $R_{\gamma}$ \\
\hline
 MARK-II~ \cite{r:MARKII-81}   & $0.041 \pm 0.008$ ($z_\gamma>0.6$ only) \\
   & $0.146 \pm 0.028$ (all $z_\gamma$, estimated) \\ \hline
 This measurement   & \Rg\ \\
\hline \hline
\end{tabular}
\end{center}
\caption{Comparison with previous experiment. MARK-II errors are
total. CLEO-c 
errors are statistical, systematic and
the uncertainty in the extrapolation to zero 
direct photon momentum. \label{tab:worldresults}}
\end{table}

\section{Implications for $\alpha_s$}
\begin{figure}[htpb]
\centerline{\includegraphics[height=9cm, width=16cm]{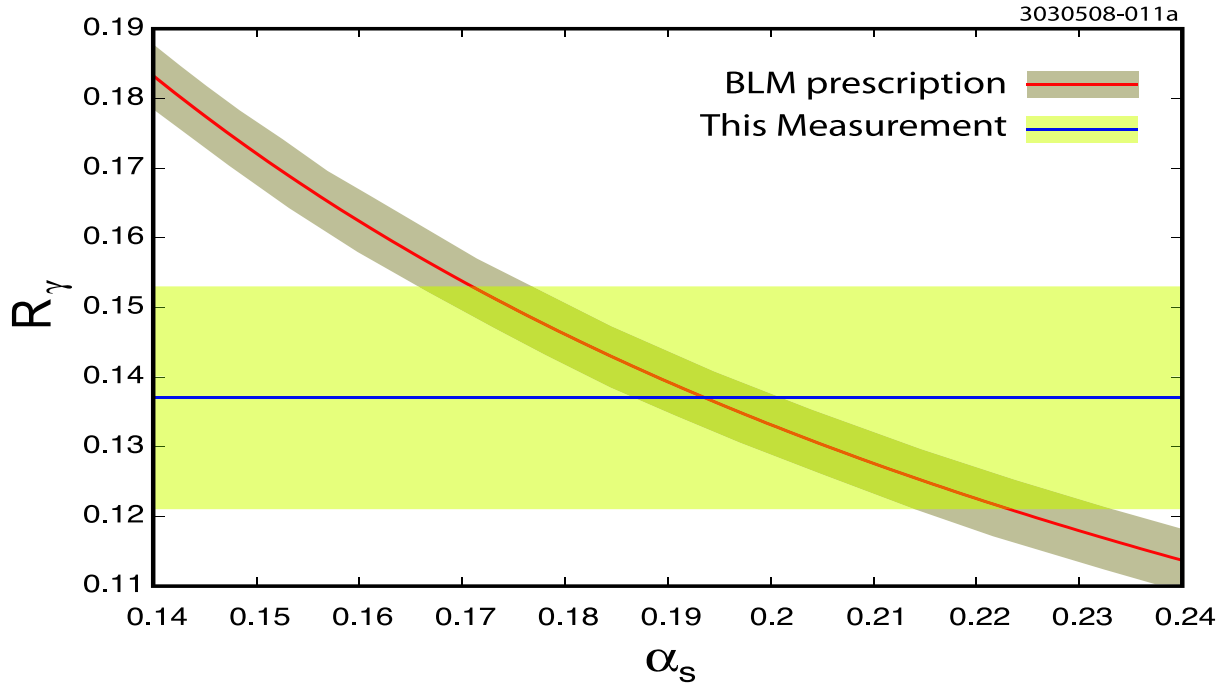}}
\caption{Horizontal band indicates $R_\gamma$
central value and experimental uncertainties; curve corresponds to
predicted $\alpha_s$ band (including theoretical errors),
using the BLM prescription for $Q^2=0.157M_{q{\overline q}}$.}
\label{bg}
\end{figure}

Although the large relativistic corrections may render such
an estimate unreliable, we can, nevertheless, calculate the value of
$\alpha_s$ implied by our $R_\gamma$ measurement, as shown in Figure \ref{bg}.
Voloshin, in his recent review \cite{Voloshin}, estimates
an expected branching fraction ${\mathcal B}(J/\psi\to gg\gamma$)=6.7\%,
using $\alpha_s(m_c)$=0.19 and the known value of $\Gamma_{ee}(J/\psi)$.
We can translate our value of $R_\gamma$ into ${\mathcal B}$ by correcting
for the non-$ggg$ decays of the $J/\psi$ (65.5\%), to obtain 
${\cal B}(J/\psi\to gg\gamma)=9.2\pm1.0$\%, considerably higher than
Voloshin's estimate. We note that the earlier MARKII result, extrapolated
to the full kinematic regime is also somewhat larger than
Voloshin's estimate.

\section{Summary}
We have extracted the direct photon energy spectrum from
$J/\psi$ decays based on a two-dimensional fit procedure.
Normalized to the dominant 3-gluon mode of the $J/\psi$, and including
two-body radiative decays,
we obtain \Rg.
Although consistent with the one previous measurement, our
direct photon yield is somewhat higher than that
expected by a simple extrapolation from results at the
$\Upsilon$(1S) ($R_\gamma=2.78\pm0.08$\%, averaged over all previous 
measurements), taking into account the variation in
$\alpha_s(Q^2)$.

\section{Acknowledgments}
We gratefully acknowledge the effort of the CESR staff
in providing us with excellent luminosity and running conditions.
Shawn Henderson wrote initial verions of the computer codes used in this
data analysis.
We thank Xavier Garcia i Tormo and Joan Soto for illuminating discussions.
D.~Cronin-Hennessy and A.~Ryd thank the A.P.~Sloan Foundation.
This work was supported by the National Science Foundation,
the U.S. Department of Energy,
the Natural Sciences and Engineering Research Council of Canada, and
the U.K. Science and Technology Facilities Council.

\end{document}